\documentclass[11pt,a4paper]{article}
\usepackage{mathrsfs, amsfonts, amssymb, amsmath, amsbsy, amscd, hyperref, graphicx}

\def\be{\begin{align}}
\def\bex{\begin{align*}}
\def\bu{$\bullet$\ \ }

\begin{document}

\title{\bf\Large Interfacial Numerical Dispersion and New Conformal FDTD Method}
\author{ Axman Fisher \\ \small  yjliu@fudan.edu.cn}
\date{}\maketitle

\abstract{This article shows the interfacial relation in
electrodynamics shall be corrected in discrete grid form which can
be seen as certain numerical dispersion beyond the usual bulk type.
Furthermore we construct a lossy conductor model to illustrate how
to simulate more general materials other than traditional PEC or
simple dielectrics, by a new conformal FDTD method which main
considers the effects of penetrative depth and the distribution of
free bulk electric charge and current.}

\tableofcontents{}\newpage

\section{Introduction}
Bulk dynamics and boundary conditions are two essential ingredients
for solving a concrete physical problem. In computational
electromagnetic, it is well known finite difference form of wave
equation in bulk leads to numerical dispersion which is measured by
the factor $\frac{\sin{x}}{x}$, where $x$ is proportional to the
grid size. If the grid length is smaller than $\frac{\lambda}{12}$,
then the dispersive factor approximates to 1 which means the error
due to this numerical dispersion can always be ignored in problem.
However, beyond this traditional bulk numerical dispersion (BND), we
find discrete process in simulation as in Finite-Difference
Time-Domain (FDTD) approach also generates certain unphysical
anti-dispersion effects on the boundary which can be defined as the
interfacial numerical dispersion (IND). If we ignore this effect, we
shall get the staircase approximation grid representation of the
curved boundary, which is just one of the shortcomings of
traditional FDTD method. Instead, we find interfacial relation in
usual electrodynamics must be corrected in grid variable form due to
IND, and some treatments in conformal FDTD method before as in
perfect electric conductor (PEC) or dielectric materials can be seen
as special cases of the new relation.

Traditional conformal FDTD (CFDTD) often deal with PEC [1-10] and
simple dielectrics [11-15]. In former case, the electric field in
material can be ignored due to the penetrative depth is far less
than one grid length due to the lossy parameter
$p=\frac{\sigma}{\epsilon \omega}\gg 1$ and in later case simple
denote near lossless, i.e. $p=\frac{\sigma}{\epsilon \omega}\ll 1$.

For general dispersive and lossy media like Lorentz model, the
related Conformal FDTD approach is difficult due to the
permittivities in the interfacial relation involve complex
dependence of the frequency and in corresponding time-domain
algorithm this shall turns to be involve with high order time
derivatives.

However, for a large class of lossy dielectrics, whose penetrative
depth $\delta$ is between PEC and perfect electric insulator (PEI),
 more exactly in which the lossy parameter $p=\frac{\sigma}{\epsilon
\omega}\sim 1$ or the $\delta$ is about several grids' length, if we
can ignore the influence of the resonance frequency in the problem
or the involved electromagnetical frequency is far away from the
material resonance frequencies, we can use certain conductivity
model as the dispersive model and get a (theoretical) more effective
CFDTD method called lossy conductor CFDTD (LCCFDTD), which consider
the effects of IND, surface/volume free electric current/charge,
penetrative depth and conformal angle (defined as the angle between
the cross-section of interface and the grid line) in more physical
way. Especially our method is more suitable for the FDTD simulation
of the problems like the design of microwave dark chamber with
strong absorber materials.

\section{Grid representation of interfacial relation}

Consider the 2-dimensional TE case below without loss of generality.
First we define the angle $\theta$ between the tangent line at the
interfacial point and the grid line as the conformal angle. The grid
line at the boundary is divided into two parts on which two electric
fields is uniform distributed as $E$ in air and $E'$ in materia.
General case of arbitrary two medias is similar.

For PEC model, the inside field can be assumed to be vanished. But
for general materials or even as in nondispersive, lossless
dielectrics, to obtain a conformal algorithm instead of naive
staircase approximation, we have to discuss the interfacial relation
between $E$ and $E'$. In continuous electrodynamics theory we have
following boundary relations:
\begin{align}
   E_\shortparallel&=E'_\shortparallel \\
   H_\shortparallel&=H'_\shortparallel+\alpha \\
  D_\bot&=D'_\bot+\rho \\
   B_\bot&=B'_\bot
\end{align}
where $\alpha$ is line density of free surface electrical current
and $\rho$ is surface density of free electrical charge. Here we
only consider the simplest case in which $\alpha,\ \rho$ vanish and
permittivity is a real const. In next section we may show how to
deal with nonzero $\alpha,\ \rho$ and complex permittivity for a
more general model of materials.

As we know, for certain frequency only two of above interfacial
relations are independent and here we use (1) (3), i.e. $$
E_\shortparallel=E'_\shortparallel,\ \ \ E_\bot=\varepsilon_r
E'_\bot,\ \  \varepsilon_r=\varepsilon/\varepsilon_0.$$ This
interfacial relation indicate $E,\ E'$ can't be along same direction
through the interface, i.e. the refraction phenomena exist or for
general case if the permittivity as function of the frequency,
 electromagnetical waves of different frequency shall refract to
different directions, i.e. dispersive phenomena appears. However, in
discrete grid simulation, $E,\ E'$ have to be along same grid line
and this in fact introduces extra error. Like the discrete treatment
for wave equation in bulk will produces numerical dispersion, the
effect of this elimination of physical refraction or dispersion on
interface due ro simulation method as FDTD can be seen as certain
interfacial numerical dispersion (IND). Hence we should take one of
two fields on the same boundary grid line as projection of the
refraction field along the grid line, take $E$ for instance and the
relation between $E,\ E'$ give the grid representation of physical
interfacial relation (1), (3).

\includegraphics[width=1.7in,height=1.6in]{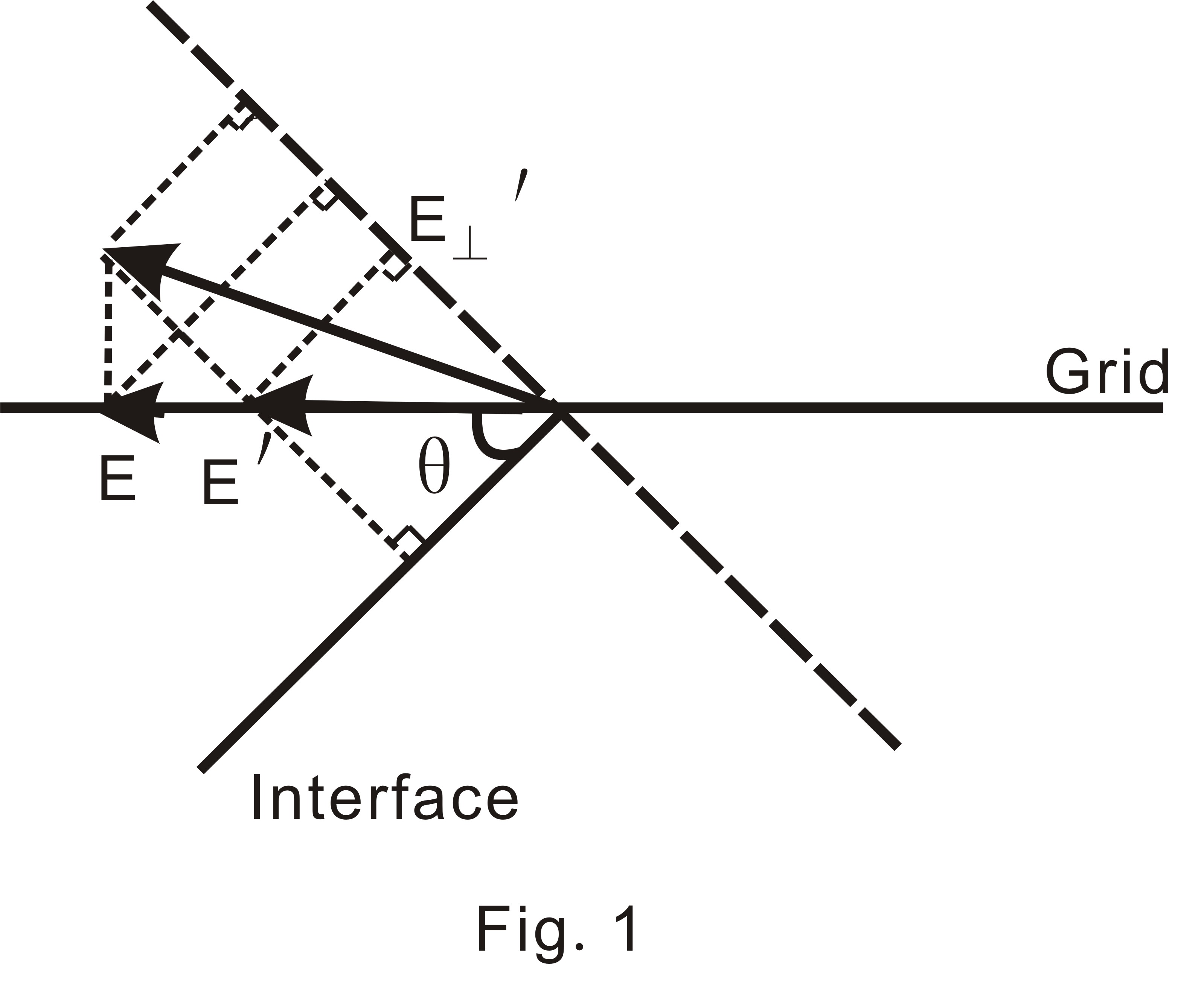}
\includegraphics[width=2.0in,height=1.9in]{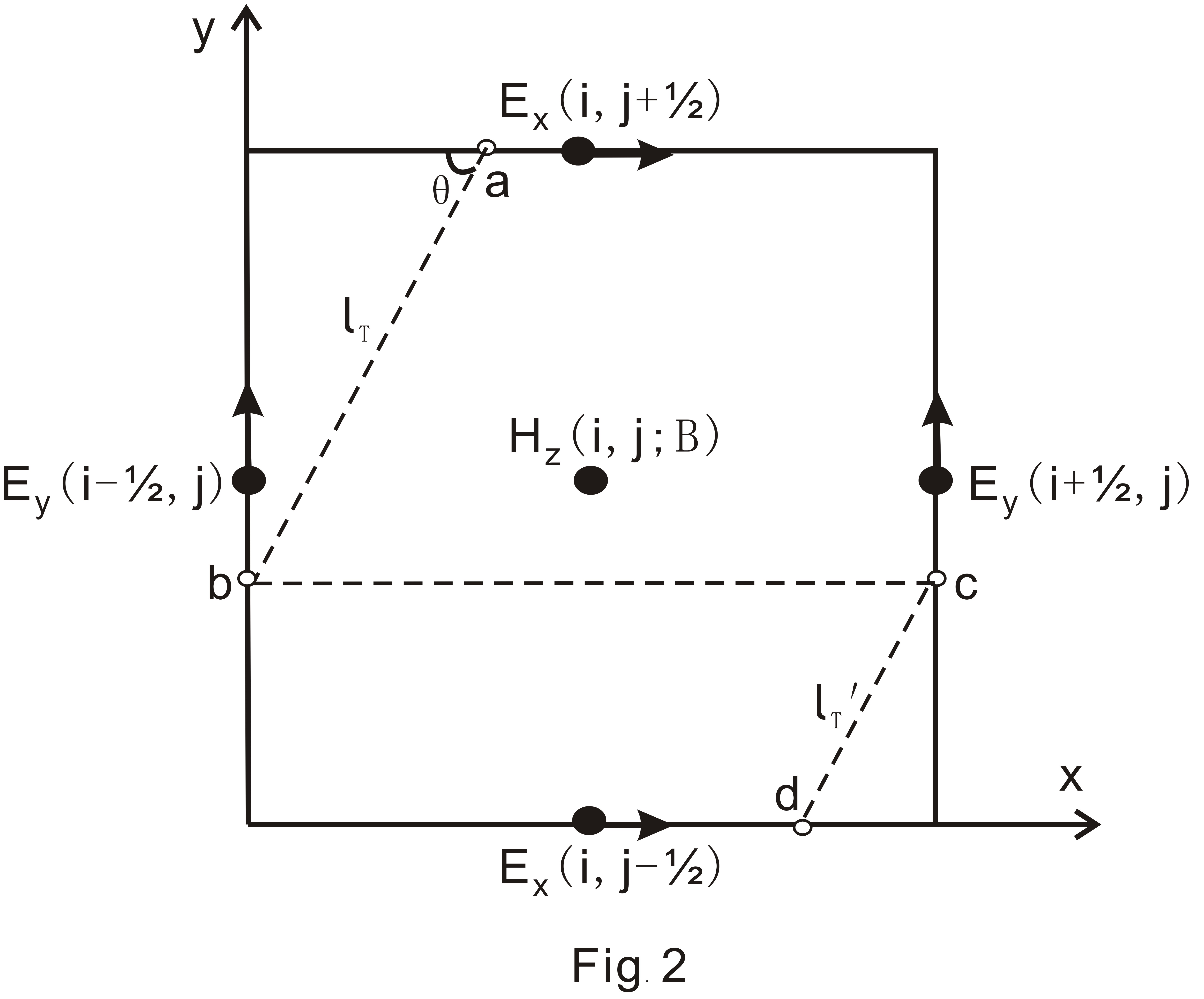}\\

By a simple geometrical analysis indicated by figure 1 above, we
find this grid interfacial relation,\be
E=(\cos^2\theta+\varepsilon_r\sin^2\theta)E'=\varepsilon_r(\theta)E'.
\end{align}
It's direct to see several special correct results. First take
$\varepsilon_r=1$ i.e. for grid in same materia, $E=E'$ as our basic
assumption in simulation. Second, consider $\theta=0$, i.e. the
coincident case of grid choice, $E=E'$, as same as traditional FDTD.
Third, if $\theta=\frac{\pi}{2}$, as the translation case for
coincident grid choice, we get $E=\varepsilon_r E'$ or equivalently
as $E_\shortparallel=E'_\shortparallel=0,\ \ D_\bot=D'_\bot$.

\section{Lossy conductor model and conformal FDTD}
 To deal with more general materials as needed in most applications, we
 construct a lossy conductor model and for
simplicity, we assume the boundary of materia is flat, i.e the
tangent line at the interfacial point is coincident with the
boundary. New model main considers a reasonable generalization of
the surface distribution of free electric charge and current in well
conductor model (WEC) via a special discrete treatment.

First we introduce a kind of lossy conductor conformal grid (LCCG),
and take a 2-dimensional cross-section of a rotated planar interface
 for example as figure 2 shows. Except usual field nodes, this grid
have four material points $a, b, c, d$ and a new grid parameter $B$
called grid penetrative number. For the grid outside of the material
$B=0$ and on the geometrical boundary of media $B=1$. For the grids
in the material, $B=\min([D_i/ds])+1$, where $D_i$ is the distance
between the grid's center node and the tangent line of interface,
$[x]$ is the maximal integer less than $x$, $ds$ denote the size of
grid. In fact we can take its relative coordinate of the nearest
interfacial grid as its $B$ value.

Only two material points in a grid are independent which are
determined by the dividing character of the boundary grids, while
other two points if needed can be obtained via translation of the
line which connects the two independent points as figure 2 shows. In
fact like traditional subgrid method which introduces four field
nodes in a grid to construct the relative small grids, our new
conformal method also consider new sub nodes which describe the
geometry of material boundary and this process don't introduces new
field nodes to increase the complexity of computation.

The start point of conformal method is the integrated form of
Maxwell equation below:\be \frac{\partial}{\partial t} \int_S \mu
H\cdot dS=-\oint_{\partial S} E \cdot dl.
\end{align}
For the case of figure 2, by discretion of above equation we have
iterated equation for magnetic field $H$, \bex
H_z^{n+\frac{1}{2}}(i,j)=&H_z^{n-\frac{1}{2}}(i,j)+\frac{\triangle
t}{\mu S(i,j)}[E_x^n
(i,j+\frac{1}{2})l_x(i,j)+E_x^{'n}(i,j+\frac{1}{2})l'_x (i,j)\\&-
E_x^n(i,j-\frac{1}{2})l_x(i,j-1)-E_x^{'n}(i,j-\frac{1}{2})l'_x(i,j-1)\\&+
E_y^n(i-\frac{1}{2},j)l_y(i-1,j)+E_y^{'n}(i-\frac{1}{2},j)l'_y(i-1,j)\\&-
E_y^n(i+\frac{1}{2},j)l_y(i,j)-E_y^{'n}(i+\frac{1}{2},j)l'_y(i,j)]
\end{align*}
Note we can just take some terms in above formula to be zero for
other cases of division. In the situation of figure 2, we also have
some constraints on the length variables due to two independent
materia points in a grid, \bex l'_y(i,j)=l'_y(i-1,j), \
l'_x(i,j-1)=l_x(i,j)\cdot \frac{l'_y(i,j)}{l_y(i-1,j)},\\
l_x+l'_x=\Delta x, \ l_y+l'_y=\Delta y; \
\frac{l_T(i,j)}{l'_T(i,j)}=\frac{l_y(i,j)}{l'_y(i,j)}. \end{align*}
It's easy to see that the usual PEC conformal algorithm is just as
the special case for which $S(i,j)$ denote the area of the outside
cell and \bex E_x(i, j+\frac{1}{2})=E'_x(i, j+\frac{1}{2}),\
E_y(i-\frac{1}{2}, j)=E'_y(i-\frac{1}{2}, j);\\  E'_x(i,
j-\frac{1}{2})=0,\ \ E'_y(i+\frac{1}{2}, j)=0.\end{align*}

By the symmetry of electric grids and magnetic grids in FDTD method,
it's not necessary to do conformal treatment for electric grids,
i.e. the iteration of electrical field is same as in traditional
FDTD which is just by discretion of another Maxwell integrated
equation:
$$ \oint_{\partial S}H\cdot dl=\frac{\partial}{\partial t}\int_S
D\cdot ds+ \int_S J\cdot ds.$$ Note in our lossy conductor model,
the conductivity $\sigma$ is not zero which require the term of free
electric current.

To deal with the effects of the free electric charge and penetrative
depth, we take a special discrete treatment for the distribution of
free bulk electric current in our grid system, which assumes the
free electric charge only distribute uniformly on the transverse
grid line $l_T$ and the electric current just conduct along the
perpendicular direction of the grid plane. Thus no electric charge
and current on the grid line in our model and especially the
transverse line $l_T$ in one grid just denote the unit transverse
line which appears in the definition of free surface electric
current $\alpha$ in (2).

By a simple deduction like for the lossless model, we can get the
generalization of formula (5) in lossy conductor model, \bex
E&=(\cos^2\theta+(\varepsilon_{r}-\frac{\sigma_r}{j\omega}+\frac{|\alpha|}{\varepsilon_0 v E'_\bot}) \sin^2\theta)E'   \\
&=(\cos^2\theta+\bar{\varepsilon}_{rc}\sin^2\theta)E'=\bar{\varepsilon}_{rc}(\theta)E'.
\end{align*}
where $|\alpha|=\sigma E' l_T$ as the electric current through the
unit transverse line $l_T$ and $v$ is the velocity of the
electromagnetical wave in the material and clearly
$v=\frac{c}{\Re(n_r)}=\frac{c}{\sqrt{\varepsilon_r\mu_r}L}$, where
the refraction lossy parameter $L=\sqrt{\frac{1+\sqrt{1+p^2}}{2}},\
\ p=\frac{\sigma_r}{\varepsilon_r \omega},\
\sigma_r=\sigma/\varepsilon_0$. Finally we obtain the grid
interfacial relation of lossy conductor model: \be
E=(\cos^2\theta+(\varepsilon_{r}-\frac{\sigma_r}{j\omega}+\frac{l_T\sigma_r\sqrt{\varepsilon_r\mu_r}L}{c\sin\theta})
 \sin^2\theta)E'  \end{align}

To apply above formula in conformal FDTD scheme, we have to make
some approximations and furthermore turn it to the iteration
relation in
time domain.\\

\bu  Half order and IND corrected approximation (HIND): $l_T=0$
(ignore the distribution of free electrical current) \be
E&=(\cos^2\theta+(\varepsilon_{r}-\frac{\sigma_r}{j\omega})
 \sin^2\theta)E'\\
\frac{\partial E}{\partial t}&=(\cos^2\theta+\varepsilon_{r}
 \sin^2\theta)\frac{\partial E'}{\partial
 t}-\sigma_r\sin^2\theta\cdot E'.
 \end{align}

\bu  First order and IND corrected approximation (FIND): \\ $l_T\ne
0,\ \ L\simeq 1$ \be
E&=(\cos^2\theta+(\varepsilon_{r}-\frac{\sigma_r}{j\omega}+\frac{l_T\sigma_r\sqrt{\varepsilon_r\mu_r}}{c\sin\theta})
 \sin^2\theta)E'\\
 \frac{\partial E}{\partial
 t}&=(\cos^2\theta+(\varepsilon_{r}+\frac{l_T\sigma_r\sqrt{\varepsilon_r\mu_r}}{c\sin\theta})
 \sin^2\theta)\frac{\partial E'}{\partial
 t}-\sigma_r\sin^2\theta\cdot E'.
\end{align}

\bu  Second order and IND corrected approximation (SIND):\\  $l_T\ne
0,\ \ L\simeq 1+\frac{p^2}{8}$ \be
E&=(\cos^2\theta+(\varepsilon_{r}-\frac{\sigma_r}{j\omega}+\frac{l_T\sigma_r\sqrt{\varepsilon_r\mu_r}}{c\sin\theta}(1+
\frac{\sigma_r^2}{8\varepsilon_r^2 \omega^2}))
 \sin^2\theta)E'\\
 \frac{\partial^2 E}{\partial
 t^2}&=(\cos^2\theta+(\varepsilon_{r}+\frac{l_T\sigma_r\sqrt{\varepsilon_r\mu_r}}{c\sin\theta})
 \sin^2\theta)\frac{\partial^2 E'}{\partial
 t^2}-\sigma_r\sin^2\theta\cdot \frac{\partial E'}{\partial t}\nonumber \\&\ \  \ -
 \frac{l_T\sigma_r^3\sqrt{\varepsilon_r\mu_r}}{8 c \varepsilon_r^2 }
 \sin\theta E'.
\end{align}

\bu Geometric boundary conformal FDTD:

Only consider above conformal treatment for the geometrical boundary
grid in which $B=1$. Thus in the iteration equation of magnetic
field we have
$$ E_x(i,j-\frac{1}{2})=E'_x(i,j-\frac{1}{2}),\ \
E_y(i+\frac{1}{2},j)=E'_y(i+\frac{1}{2},j)$$ and
$E_x(i,j+\frac{1}{2}),\ E_y(i-\frac{1}{2},j)$ can be determined by
$E'_x(i,j+\frac{1}{2}),\ E'_y(i-\frac{1}{2},j)$ via the discrete
form of the equation (9), (11) and (13). For $B>1$, we set $E=E'$
which means we have ignored the effect of penetrative depth in this
scheme.\\

\bu Penetrative boundary conformal FDTD:

Consider above conformal treatment for the penetrative boundary grid
in which $1 \leq B\leq [\frac{\delta}{\Delta l}],\ \ \Delta l=\min
(\Delta x,\ \Delta y )$. For such grids we get all $E$ from $E'$ via
discrete form of (9), (11) and (13). The grid with $B=
[\frac{\delta}{\Delta l}]+1$ will be treated as geometric boundary
grid while for $B>[\frac{\delta}{\Delta l}]+1$ we just set $E=E'$.
Especially for PEC case, $[\frac{\delta}{\Delta l}]=0$ and $E'=0$ as
in traditional way.

Although the penetrative depth $\delta$ is an important physical
quantity which can be determined by the electromagnetic parameters
of the material exactly, we can take it as an adjustable parameter
called lossy depth $\delta_l$ for practical use. It can be set by
the precision and cost of computation in a problem and penetrative
depth $\delta$ could just be its default value.

\section{Discussion}

We know electrodynamics manifest Lorentz symmetry or Maxwell
equations in continuous spacetime is covariant under the translation
and rotation, while in discrete grid spacetime such Poincare
symmetry is broken. For flat boundary material, we find formula (5)
considers the exact effect of conformal angle in the interfacial
relation such that the simulation physical result is independent of
grid system in the sense of global translation and rotation, or in
other words for flat boundary case the shape of material in
simulation is invariant of grid translation and rotation which is
more physical unlike in traditional FDTD. In this manner we could
say our IND corrected conformal FDTD have restored certain
3-dimensional space symmetry and especially it manifests the
consistence between the conformal and nonconformal algorithm.

There are several directions to generalize above section's result.

First, if the boundary is curved, the tangent line at the
interfacial point shall not be  along $l_T$. One simple way to deal
with this case is to define the angle between the  tangent line and
transversal line $l_T$ as relative conformal angle $\theta_r$ and
the conformal angle still to be the one between $l_T$ and the grid
line. If $\theta_r\leq\frac{\theta}{2}$, such boundary's treatment
can be approximated by transversal line boundary as done in
previous, while $\theta_r>\frac{\theta}{2}$, we'd better approximate
the boundary in this direction by the complete grid line, i.e.
$E=E'$.

Second, our LCCFDTD method only considers the electrical loss, it
seems the magnetic loss and some special anisotropy or nonlinear or
dispersive medias also can be treated in similar way [16]. It may
also have applications in other numerical computational areas.

Finally, we may consider the stability problem as in traditional
conformal FDTD method [17-18] and  an improved absorb boundary may
be obtained by the combination of a large lossy conductor material
and the ideas of PML. Of course all above discussions are main in
theoretical manner and it can be verified by some numerical
experiments. The results may be similar with [14], i.e. the
simulation of rotated problem should has good
agreement with the coincidence problem as reference solution.\\


\begin{thebibliography}{aa}


\bibitem{1} Thomas G. Jurgens, Allen Taflove et al., ¡°Finite-Difference
Time-Domain Modeling of Curved Surfaces¡±, IEEE Trans.  Antennas
Propagat.,  40(1992) 357.

\bibitem{2} T. Jurgens and A. Taflove, "Three-dimensional contour FDTD modeling
of scattering from single and multiple bodies," IEEE Trans. Antennas
and Prop., vol. 41, pp. 1803-1708, Dec. 1993.

\bibitem{3} S.Dey, R.Mittra, S.Chebolu, ¡°A technique for implementing the
FDTD algorithm on a nonorthogonal grid¡±, IEEE Trans. Antennas
Propagat., 40(1992)357.

\bibitem{4} S. Dey and R. Mittra, "A locally conformal FDTD
algorithm for modeling 3D PEC objects," IEEE Microwave and Guided
Wave Letters, vol. 7, pp. 273-275, Sept. 1997.

\bibitem{5} C. J. Railton and J. Schneider, "An analytical and numerical
analysis of several locally conformal FDTD schemes," IEEE Trans. On
Microwave Theory and Techniques, vol. 47, pp. 56-66, Jan. 1999.

\bibitem{6} Mark W.Steeds, et al., ¡°A Comparison of Two Conformal Methods
for FDTD Modeling¡±, IEEE Trans. Electr. Compat., 38(1996)181.

\bibitem{7} Y. Hao and C.J. Railton, Analyzing electromagnetic structures with
curved boundaries on Cartesian FDTD meshes, IEEE Trans Microwave
Theory Tech 46(1998), 82-88.

\bibitem{8} I.A.Zagorodnov, R.Schuhmann, ¡°A uniformaly stable conformal
FDTD-method in Cartesian grids¡±, Int. J. Numer. Model.,
16(2003)127.

\bibitem{9} I.A.Zagorodnov, R.Schuhmann, ¡°Conformal FDTD-methods to avoid
time step reduction with and without cell enlargement¡±, J. Comput.
Phys.,  225(2007)1493.

\bibitem{10} Tian Xiao and Qing Huo Liu, ¡°A 3-D Enlarged Cell Technique (ECT)
for the Conformal FDTD Method¡±, IEEE Trans. Antenn. Prop.,
225(2007)1493.

\bibitem{11} R. D. Meade, et al., "Accurate theoretical analysis of photonic
band-ap materials," Phys. Rev. B, vol. 48, pp. 8434-8437, 1993.

\bibitem{12} N. Kaneda, B. Houshmand, and T. Itoh, FDTD analysis of dielectric
resonators with curved surfaces, IEEE Trans Microwave The- ory
Technol 45 (1997), 1645-1649.

\bibitem{13}  S. Dey and R. Mittra, A locally
conformal finite difference time domain technique for modeling
arbitrary shaped objects, IEEE Antennas Propagat Soc Int Symp, June
1998, vol. 1, pp. 584-587.

\bibitem{14} J.-Y. Lee and N.-H. Myung,
"Locally tensor conformal FDTD method for arbitrary dielectric
surfaces," Microwave and Optical Technology Letters, vol. 23, No. 4,
pp. 245-249, Nov. 20, 1999.

\bibitem{15} Nadobny J, Sullivan D, Wlodarczyk W, et al.,¡±A 3-D tensor
FDTD formulation for treatment of sloped interfaces in electrically
inhomogeneous media,¡± IEEE T. Antennas and Propagation, vol. 51
(8), pp. 1760-1770, August 2003.

\bibitem{16} Luebbers R., Lossy dielectrics in FDTD, IEEE transactions on antennas and propagation, 1993.

\bibitem{17} Pereda J.A., Garcia O.,  Vegas A.,  Prieto A., Numerical
dispersion and stability analysis of the FDTD technique in lossy
dielectrics,  Microwave and Guided Wave Letters, IEEE, Jul. 1998.

\bibitem{18} A.C. Cangellaris and D.B. Wright, Analysis of the numerical error
caused by the stair-stepped approximation of a conduction boundary
in FDTD simulations of electromagnetic phenomena, IEEE Trans
Antennas Propagat 39(1991), 1518-1525.

\end{thebibliography}
\end{document}